\documentstyle{article}
\begin{document}

\title{The shear dynamics in Bianchi I  cosmological model
on the brane}

\author{A.V.Toporensky}
\date{}
\maketitle
\hspace{8mm}{\em Sternberg Astronomical Institute,
 Moscow 119899, Russia}\\
\begin{center}
 Electronic mail: {\tt lesha@sai.msu.ru}
\end{center}

\begin{abstract}
The shear dynamics in Bianchi I cosmological model 
on the brane with a perfect fluid (the equation of state is $p=(\gamma-1)
\mu$) is studied. It is shown
that for $1 < \gamma < 2$ 
the shear parameter has maximum at some moment during a transition period
from nonstandard
 to standard  cosmology.
An exact formula for the matter density $\mu$ in the epoch of maximum
shear parameter as a function of 
 the equation of state is obtained.
\end{abstract}

In recent several years a lot of effort has been done on the idea that
our Universe is a boundary of a space-time manifold with a larger
number of dimensions \cite{brane}.
The cosmological evolution of a three-brane in five dimensional
space-time become the matter of intense investigations. The corrections
to Einstein equations on the brane arising from the influence of a bulk
geometry on the brane were obtained  \cite{Maeda}.
The corresponding cosmological dynamics was studied in detail for a FRW Universe.
It was shown that the dynamics of the early brane Universe can differ
significantly from the standard scenario \cite{Binetruy}.
On the other hand, the standard FRW cosmology for the late Universe
(when the matter density on the brane $\mu$ become negligible
in comparison with the brane tension $\lambda$) was recovered \cite{B-C}.

Later on
this analisys has been done for an anisotropic Bianchi I metric on the brane.
The matter source was chosen in the form of a massive scalar field
\cite{Varun} or perfect fluid with the equation of state $p=(\gamma-1) \mu$
\cite{Campos} (see also \cite{Chen} where some exact solutions were found).
One of the most interesting results obtained in this way is that the initial
singularity, being anisotropic in ordinary cosmology (with the exception
of the maximally stiff fluid case \cite{Ellis}), becomes isotropic
in the brane cosmology for $\gamma  \in (1,2)$\cite{Varun,Campos}.
Since during the  evolution of the brane Universe its dynamics approaches
the standard one,
later stages of the
Universe are isotropic for $\gamma < 2$.
So, the intermediate stage of a relatively high anisotropy can exist if
the matter filling the Universe has a positive pressure. Intuitively we
could expect that the anisotropy maximum is reached sometimes in the transitional
period when the matter density is comparable with the brane tension.
In this paper we calculate the ratio $\mu/\lambda$ during the maximum
anisotropy epoch. We will see that for the radiation-dominated Universe
the above suggestion appears to be true, though for an arbitrary $\gamma$
from the interval $(1,2)$  (we study only this type of matter
in this paper) the epoch of maximal anisotropy can differ significantly
from the period when $\mu \sim \lambda$. To conclude this introductory
part we remind that current experimental limit on the brane tension is
$ \lambda > (100 \mbox{GeV})^4$ \cite{Maartens}.

The field equations on the brane are \cite{Maartens}
\begin{equation}
G_{\mu \nu}=-\Lambda g_{\mu \nu} + \kappa^2 T_{\mu \nu} +
\tilde \kappa^4 S_{\mu \nu} - {\cal E}_{\mu \nu}.
\end{equation}
Here $\tilde \kappa$ is the 5-dimensional gravitational constant,
$\kappa$ is the effective 4-dimensional gravitation constant on the brane.
Below we assume that
the effective cosmological constant on the brane $\Lambda = 0$. Bulk
corrections to the Einstein equations on the brane are of two forms:
there are quadratic energy-momentum corrections via the tensor $S_{\mu \nu}$
and nonlocal effects from the free gravitational field in the bulk.
The energy-momentum corrections are given by
\begin{equation}
S_{\mu \nu} = \frac{1}{12}T T_{\mu \nu}-\frac{1}{4}
T_{\mu}{}^{\delta} T_{\mu \delta}+\frac{1}{24}g_{\mu \nu}\left[3T^{\delta
\rho}
T_{\delta \rho}-T^2\right] ,
\end{equation}
where $T\equiv T_{\mu}{}^{\mu}$.
If we define $u^{\mu}$ as the 4-velocity comoving with matter, the nonlocal
term has the form
\begin{equation}
{\cal E}_{\mu\nu}={-6\over\kappa^2\lambda}\left[{\cal
U}\left(u_\mu u_\nu+{\textstyle {1\over3}} h_{\mu\nu}\right)+{\cal
P}_{\mu\nu}+{\cal Q}_{\mu}u_{\nu}+{\cal Q}_{\nu}u_{\mu}\right],
\end{equation}
where $h_{\mu \nu}=g_{\mu \nu} + u_{\mu} u_{\nu}$. Here
$$
{\cal U}=-\frac{1}{6}\kappa^2 \lambda\, {\cal
E}_{\mu\nu}u^\mu u^\nu
$$
is an effective nonlocal energy on the brane,
$$
{\cal P}_{\mu\nu}=-\frac{1}{6}\kappa^2 \lambda\left[
h_\mu{}^\alpha h_\nu{}^\beta-\frac{1}{3}h^{\alpha\beta}
h_{\mu\nu}\right] {\cal E}_{\alpha\beta}
$$
is an effective nonlocal anisotropic stress and
$$
{\cal Q}_\mu =\frac{1}{6}\kappa^2 \lambda\,
h_\mu{}^\alpha {\cal E}_{\alpha\beta}u^\beta
$$
is an effective nonlocal energy flux.

There are also conservation equations for the brane energy-momentum
tensor $\bigtriangledown^{\mu}T_{\mu \nu}=0$.
Full description of dynamical and conservation equations see in
\cite{Maartens}.

We will study Bianchi I model on the brane with the metric
\begin{equation}
ds^2 = -dt^2 + a_i^2(t) (dx^i)^2,
\end{equation}

The conservation equations in this case reduce to
\begin{eqnarray}
&&\dot{\mu}+\Theta(\mu+p)=0,\\ && \dot{\cal
U}+\frac{4}{3}\Theta{\cal U}+\sigma^{\mu\nu}{\cal
P}_{\mu\nu}  =0 , \\ && D^\nu{\cal P}_{\mu\nu}
=0,
\end{eqnarray}
where $\Theta$ is the volume
expansion rate, and $\sigma_{\mu\nu}$ is the shear.
The nonlocal energy flux ${\cal Q}$ vanishes in this case
identically and we have no evolution
equation for ${\cal P}_{\mu \nu}$ which is a bulk degree
of freedom and can not be predicted
from the brane \cite{Varun,Maartens}.   In terms of the mean scale
factor $a=(a_1 a_2 a_3)^{1/3}$ the expansion rate $\Theta = 3H = 3\dot a/a$.

The Raychaudhuri equation on the brane is \cite{Varun}
\begin{equation}
\dot{\Theta}+\frac{1}{3}\Theta^2
+\sigma^{\mu\nu}\sigma_{\mu\nu}
+\frac{1}{2}\kappa^2(\mu + 3p)=
-\frac{1}{2} \kappa^2 (2\mu+3p){\mu\over\lambda} -{6
{\cal U}\over\kappa^2\lambda}
\end{equation}

and Gauss-Codazzi equations are \cite{Varun}
\begin{eqnarray}
\dot{\sigma}_{\mu\nu}+\Theta\sigma_{\mu\nu}&=&{6
\over\kappa^2\lambda}{\cal P}_{\mu\nu},\\
-\frac{2}{3}\Theta^2 +\sigma^{\mu\nu}\sigma_{\mu\nu}
+2\kappa^2\mu &=& -\kappa^2{\mu^2\over\lambda} -{12 {\cal
U}\over\kappa^2\lambda}.
\end{eqnarray}

Due to the presence of the nonlocal stress ${\cal P}_{\mu \nu}$ this system
of equations is not closed. In previous papers
\cite{Varun, Campos} this difficulty was avoided by choosing ${\cal U}=0$.
This condition, however, is sufficient, but not necessary for closing
 the system. The full situation can be described as follows.

The most restrictive condition is ${\cal P}_{\mu \nu}=0$. In this case it is possible
to solve the equation for the shear {\it separately} for components of
the shear tensor. The less restrictive condition 
$\sigma^{\mu \nu} {\cal P}_{\mu \nu}=0$ allows us to find the dynamics of shear scalar
$\sigma^{\mu \nu} \sigma_{\mu \nu}$ leaving components of the shear
tensor undefined.

Taking into account Eq. (6), we can see that both these assumptions
lead to Friedmann-like behavior of the "dark radiation" term
$$
{\cal U}=\frac{C}{a^4}
$$
with $C=Const$.

The condition ${\cal U}=0$ automatically leads to $\sigma^{\mu \nu}
{\cal P}_{\mu \nu}
=0$ (see Eq. (6)). However, the latter condition
 is more general. We
assume further that $\sigma^{\mu \nu} {\cal P}_{\mu \nu}=0$ is satisfied.
In the opposite case
the nonlocal anisotropic stress changes significantly the dynamical
equations for shear and "dark radiation". Though even in this case
there are some possibilities for closing the system (8)-(10) \cite{Santos}
the resulting dynamics can be very different from studied in this
paper and should be specially investigated.

It should be noted that though there are no restriction to choose
${\cal P_{\mu \nu}}$ on the brane, we do not know whether a particular
chose is consistent in a bulk. The general form of
bulk anisotropic metric is not known
yet (the first approach to this problem see in \cite{Frolov}). We
leave the problem to check that ${\cal P_{\mu \nu}}$ chosen
on the brane is consistent with the full 5-dimensional metric to
future investigations.

It is convenient to introduce the deceleration parameter $q$ as
 $$
q=-\frac{\ddot a a}{\dot a^2},
$$
or, equivalently,
\begin{equation}
\dot H=-(1+q)H^2.
\end{equation}
This equation in combination with Eq.(8), which can be written in the form
\begin{equation}
\dot H=-H^2-\frac{2}{3}\sigma^2-\frac{\kappa^2 \mu}{6}(3 \gamma-2) -
\frac{\kappa^2 \mu^2}{6 \lambda}(3 \gamma-1) - \frac{2 {\cal U}}
{\kappa^2 \lambda},
\end{equation}
gives the expression for $q$. To obtain this expression in more
convenient form we introduce dimensionless variables
\begin{equation}
\Sigma^2=\frac{\sigma^2}{3 H^2},
\end{equation}

\begin{equation}
\Omega_{\lambda}=\frac{\kappa^2 \mu^2}{6 \lambda H^2},
\end{equation}

\begin{equation}
\Omega_{\mu}=\frac{\kappa^2 \mu}{3 H^2},
\end{equation}

\begin{equation}
\Omega_{\cal U}=\frac{2 {\cal U}}{\kappa^2 \lambda H^2}.
\end{equation}

Substituting Eqs.(13)-(16) into (12) we obtain
\begin{equation}
q=2 \Sigma^2 + \frac{1}{2}(3 \gamma-2)\Omega_{\mu}+(3 \gamma-1)
\Omega_{\lambda}+\Omega_{\cal U}
\end{equation}
and the constraint equation (10) in the form
\begin{equation}
1=\Sigma^2 + \Omega_{\lambda} + \Omega_{\mu} + \Omega_{\cal U}.
\end{equation}

With a new time variable $\tau$ defined as $\frac{dt}{d\tau}=\frac{1}{H}$
the dynamical equation for the shear takes the form as in standard
Bianchi I cosmology \cite{Ellis}
\begin{equation}
(\Sigma^2)'=2(q-2)\Sigma^2
\end{equation}
where $'$ is the derivative with respect to $\tau$.
We remind, however, that unlike standard cosmology we can not determine
shear components separately unless ${\cal P}_{\mu \nu}=0$.
The Eqs. (17)-(19) take the form obtained in \cite{Campos} in the
particular case ${\cal U}=0$.
 Using (19) we can easily find that the extremum of the shear parameter
$\Sigma$ corresponds to deceleration parameter $q=2$. Expressing $\Sigma$
from (18) and substituting into (17) we find that
\begin{equation}
q=2 + \frac{3}{2} \Omega_{\mu}(\gamma-2) + 3\Omega_{\lambda}(\gamma-1) -
\Omega_{\cal U}.
\end{equation}
This leads to the following equation satisfied at the moment when $\Sigma$
reaches its extremum:
\begin{equation}
\Omega_{\cal U}=\frac{3}{2}\Omega_{\mu}(\gamma-2) + 3\Omega_{\lambda}(\gamma-1).
\end{equation}

First let us assume ${\cal U}=0$. In this case it is possible to obtaine
the result in a very compact form. The equation (21) yields now
$$
\Omega_{\lambda}(\gamma-1)=-\frac{1}{2}\Omega_{\mu}(\gamma-2)
$$
Since $\Omega_{\lambda}$ falls more rapidly than $\Omega_{\mu}$, this
extremum of shear is indeed its maximum.

Remembering the definitions of $\Omega_{\lambda}$ and $\Omega_{\mu}$
we can write that $\Omega_{\mu}/\Omega_{\lambda}=2 \lambda/\mu$.
Using this formula we finally get the ratio of matter density and
the brane tension at the moment of maximal shear parameter for a given
equation of state:
\begin{equation}
\frac{\mu}{\lambda}=\frac{2-\gamma}{\gamma-1}.
\end{equation}

For the most interesting case of radiation-dominated matter ($\gamma=4/3$)
the maximum of $\Sigma$ corresponds to $\mu=2 \lambda$.

In the case of the maximally stiff matter ($\gamma=2$) the isotropisation does not
take place even in the standard cosmology, so for $\gamma$ close to $2$
the shear parameter can reach its maximum at an arbitrary late time with an
arbitrary small
ratio $\mu/\lambda$. In the opposite case of a very low pressure the
isotropisation can begin at a very high matter density. If the pressure
is exactly equal to zero ($\gamma=1$) or it is negative ($\gamma < 1$),
the shear parameter decreases from the beginning. In this case there is no
maximum of $\Sigma$ and formula (22) become inapplicable.

If, as often done, we write the equation of state through the
parameter $w$ as $p=w\mu$, Eq. (22) takes even more simple
and transparent form

$$
\frac{\mu}{\lambda}=\frac{1-w}{w},
$$
$w \in (0,1)$.

In the case ${\cal U} \ne 0$ it is necessary to use the full form of (21).
 It can be
rewritten as a quadratic equation for $\mu/\lambda$
\begin{equation}
\frac{\mu^2}{\lambda^2}(\gamma-1) + \frac{\mu}{\lambda}(\gamma-2)
=\frac{4 {\cal U}}{\kappa^4 \lambda^2}
\end{equation}
but with righthandside decreasing with time. For
positive ${\cal U}$ (otherwise the brane Universe can recollape
without reaching the isotropic stage \cite{Santos})
Eq. (23) has only
one positive root, corresponding to maximum of the shear parameter.

  As ${\cal U}$ decreases like radiation
(${\cal U}=C/a^4$), the result obtained for radiation matter without ${\cal U}$
can be easily generalised. For radiation-dominated Universe
 it is possible to write
$$
\frac{\cal U}{\kappa^4 \lambda} = \alpha \mu
$$
with some dimensionless constant parameter $\alpha$. 
Eq.(23) gives now the matter
density at the time of shear maximum as $\mu = \lambda (2 + 12 \alpha)$. 
In general, for positive ${\cal U}$  the isotropisation
time shifts to an earlier time than for ${\cal U}=0$. In particular,
in the case of the maximally stiff fluid the shear parameter has the maximum
at
$$
\mu=\frac{2}{\kappa^2}\sqrt{\cal U}.
$$
Clearly, $\mu \to 0$ with ${\cal U} \to 0$, reflecting the fact that there is
no isotropisation for the maximally stiff fluid without "dark radiation".

We investigated the shear dynamics in Bianchi I cosmology on the brane.
Unlike ordinary General Relativity scenario (where the shear parameter
monotonically decreases from Kasner initial singularity to isotropic
future attractor unless $\gamma=2$)
this dynamics for $\gamma \in (1,2)$ describes
the evolution of the Universe from isotropic singularity to isotropic
future attractor through an intermediate anisotropic stage.
It was shown that assuming $\sigma^{\mu \nu} {\cal P}_{\mu \nu}=0$
 on the brane it is possible to find
the matter density in the epoch of maximum shear parameter analytically.
A more restrictive condition ${\cal P_{\mu \nu}}=0$ on the brane allows
us to separate the evolution of shear tensor components. An important
problem
for future development is to find 5-dimensional bulk metric for a
Bianchi I brane-world and check the consistency of these ${\cal P_{\mu
\nu}}$
forms in the bulk.

\section*{Acknowledgements}
Author is greatful to Varun Sahhi and Tarun Deep Saini
for stimulated discussions and to IUCAA, where this work was done,
for hospitality.


\begin{thebibliography}{99}
\bibitem{brane}
P.Ho$\check{\mbox{r}}$ava and E.Witten, Nucl. Phys. {\bf B460},  506  (1996);
L.Randall and R.Sundrum, Phys. Rev. Lett {\bf 83},  4690  (1999).
\bibitem{Maeda}
T.Shiromizu, K.Maeda, and M.Sasaki, Phys. Rev. D {\bf 62},  024012  (2000);
M.Sasaki, T.Shiromizu, and K.Maeda, Phys. Rev. D {\bf 62},  024008  (2000).
\bibitem{Binetruy}
P.Bin$\acute{\mbox{e}}$truy, C.Deffayet, and D.Langlois, Nucl. Phys. {\bf
  B565},  269  (2000);
\bibitem{B-C}
C.Cs$\acute{\mbox{a}}$ki, M.Graesser, C.Kolda and J.Terning,
Phys. Lett. {\bf B462}, 34 (1999);
P.Bin$\acute{\mbox{e}}$truy, C.Deffayet, U.Ellwanger, and D.Langlois, Phys.
  Lett. B {\bf 477},  285  (2000);
\bibitem{Varun} R.Maartens, V.Sahni and T.D.Saini, Phys. Rev. D {\bf 63},
063509  (2001).
\bibitem{Campos} A.Campos and C.F.Sopuerta,
"Evolution of cosmological models in the brane-world scenario",
hep-th/0101060.
\bibitem{Chen} Chiang-Mei Chen, T.Harko and M.K.Mak,
"Exact anisotropic brane cosmologies", hep-th/0103240.
\bibitem{Ellis} J.Wainwright and G.F.R.Ellis,
{\em Dynamical systems in cosmology} (Cambridge University Press,
 Cambridge, 1997);
\bibitem{Maartens}
R.Maartens, Phys. Rev. D {\bf 62},  084023  (2000);
R.Maartens, "Geometry and dynamics of the brane world", gr-qc/0101059.
\bibitem{Santos} M.G.Santos, F.Vernizzi and P.G.Ferreira,
"Isotropisation and instability of the brane", hep-ph/0103112.
\bibitem{Frolov} A.V.Frolov, "Kasner-AdS spacetime and anisotropic
brane-world cosmology", gr-qc/0102064. 
\end{thebibliography}
\end{document}